\documentclass[12pt]{article}

\newcommand{\ltwid}{\mathrel{\raise.3ex\hbox{$<$\kern-.75em\lower1ex\hbox{$\sim$
}}}}
\newcommand{\gtwid}{\mathrel{\raise.3ex\hbox{$>$\kern-.75em\lower1ex\hbox{$\sim$
}}}}
\newcommand{\s}{\scriptscriptstyle}

\usepackage{epsf}

\usepackage{scicite}

\newenvironment{sciabstract}{%
\begin{quote} \bf}
{\end{quote}}

\newcounter{lastnote}
\newenvironment{scilastnote}{%
\setcounter{lastnote}{\value{enumiv}}%
\addtocounter{lastnote}{+1}%
\begin{list}%
{\arabic{lastnote}.}
{\setlength{\leftmargin}{.22in}}
{\setlength{\labelsep}{.5em}}}
{\end{list}}

\title{A Primordial Origin of the Laplace Relation Among the Galilean
       Satellites}

\author
{S. J. Peale and Man Hoi Lee\\
\\
\normalsize{Department of Physics, University of California,}\\
\normalsize{Santa Barbara, CA 93106, USA}\\
}
\begin{document} 
\date{}

\maketitle 

\begin{sciabstract}
Understanding the origin of the orbital resonances of the Galilean satellites
of Jupiter will constrain the longevity of the extensive volcanism on Io,
may explain a liquid ocean on Europa, and may guide studies of the
dissipative properties of stars and Jupiter-like planets.
The differential migration of the newly formed Galilean satellites due 
to interactions with a circumjovian disk can lead to the
primordial formation of the Laplace relation
$\mbox{\boldmath $n_1-3n_2+2n_3=0$}$, where the 
$\mbox{\boldmath $n_i$}$ are the mean 
orbital angular velocities of Io, Europa, and Ganymede, respectively.  
This contrasts with the formation of 
the resonances by differential expansion of the orbits from
tidal torques from Jupiter.
\end{sciabstract}

\noindent
The orbital resonances among the Galilean satellites of Jupiter have led to
sustained dissipation of tidal energy to produce astounding volcanos
on Io and, probably, to maintain a liquid ocean on Europa. Understanding
the origin of these resonances will constrain the history of the
satellites and their formation scenarios. The two models
proposed are an assembly of the resonances through differential
tidal expansion of the orbits from tides raised on Jupiter
\cite{yod79,yod81} and a primordial origin of unspecified assembly and
subsequent evolution away from exact resonance
\cite{gre82,gre87}. Here we discuss a means of assembling the
resonances during the formation process in support of a primordial
origin.     

The currently observed orbital resonances at the 2:1 mean motion
commensurabilities involving 
Io-Europa and Europa-Ganymede are such that the resonance 
variables $\theta_1=\lambda_1-2\lambda_2+\varpi_1$ and
$\theta_3=\lambda_2-2\lambda_3+\varpi_2$ librate about $0^\circ$ 
and $\theta_2=\lambda_1-2\lambda_2+\varpi_2$ librates about
$180^\circ$, all with small amplitude.  Here the $\lambda_i$ are
mean orbital longitudes and the $\varpi_i$ are the longitudes of periapse
numbered consecutively for Io, Europa, and Ganymede, respectively. The
simultaneous libration of $\theta_2$ and $\theta_3$ means the Laplace
angle $\theta_5=\theta_2-\theta_3=\lambda_1-3\lambda_2+
2\lambda_3\approx 180^\circ$ 
with small libration amplitude, leading to $n_1-3n_2+2n_3=0$, where
$n_i=\dot\lambda_i$ are the mean angular velocities with the dot
indicating time differentiation. This last
libration condition is called the Laplace relation because Laplace
first understood its stability.  The geometry 
is such that conjunctions of Io and Europa occur when Io is near the
periapse of its orbit and Europa is near its apoapse, and conjunctions
of Europa and Ganymede occur when Europa is near its periapse.
Ganymede can be anywhere in its orbit at the conjunctions with Europa,
because the resonance variable $\theta_4=\lambda_2-2\lambda_3+\varpi_3$
circulates through all angles.   The existence of this group of
resonances with small amplitudes of libration implies an
assembly of originally random orbits by dissipative processes. 

A currently
popular means of resonance assembly is by differential orbital
expansion due to gravitational tides raised on Jupiter, principally by
Io, and damping of the libration amplitudes by tidal dissipation
within the satellites, mostly in Io \cite{yod79,yod81}.
Io's orbit expands more rapidly than Europa's, and capture of Io
and Europa into the 2:1 orbital resonances is certain for a reasonably
wide range of initial orbital eccentricities. Continued orbital
expansion tends to drive the pair deeper into the resonance (i.e.,
driving $n_1-2n_2$ to smaller values) and
thereby to increase the orbital eccentricities. But dissipation of tidal
energy within the satellites tends to decrease the eccentricities, so an
equilibrium can be reached where the eccentricities and the ratio of the
mean motions are almost constant as the pair of satellites
moves out together locked in the resonances  \cite{yod79,yod81}.
 Europa eventually encounters the 2:1 mean motion
commensurability with Ganymede, where capture into the Laplace
relation as the resonance variable $\theta_3$ is trapped into
libration about $0^\circ$ has a probability of about 0.9 \cite{yod79}.  
$\theta_4$ is circulating because the free eccentricity \cite{free} of
Ganymede's 
orbit exceeds the relatively small forced eccentricity in the
resonance \cite{pea99}. The librations are 
damped by the tidal dissipation in Io. Variations of this scenario have been 
proposed by Malhotra and colleagues \cite{mal91,sho97,ssm97}.
The creation of the Laplace relation by
differential tidal expansion of the orbits depends on the dissipation
function $Q_{\s J}$ of Jupiter being sufficiently small ($\sim 10^6$) for
non-negligible expansion of the orbits to occur,
and the maintenance of the current configuration in equilibrium would
require an even smaller $Q_{\s J}$ ($\sim 2 \times 10^4$).

Greenberg, motivated by the difficulty of
accounting for a sufficiently small $Q_{\s J}$ necessary for the tidal
origin or equilibrium described above and the observed positive $\dot n_1$,
proposed a primordial origin of the Laplace 
relation \cite{gre82,gre87}. He described evolution from much deeper in the 
resonances (i.e., with $n_1-2n_2$ approaching zero and thereby forcing
much larger orbital eccentricities). The path from deep resonance as
dissipation decreased the eccentricities is characterized by stable
librations of the resonance angles about centers much different from
the $0^\circ$ or $180^\circ$ currently observed. The high dissipation
rates at the high eccentricities imply rapid decay of the resonances
($<10^7$ year time scale), and the only way to preserve them for
$4.6\times 10^9$ years was to have a sufficiently low $Q_{\s
J} \sim 10^6$ to allow at least an episodic heating and cooling of Io,
albeit not as low as would be required by the equilibrium model. Now we
would be observing the system on its way out of the resonance,
which lowers the eccentricities and, hence, the heating rates. Eventually
Io cools and the dissipation function $Q_{\s I}$ of Io increases to the
point at which the Jupiter torque ``wins,'' pushing the system back to higher
eccentricities to repeat the cycle \cite{gre82,oja86}.  It is not certain
that Io could cool rapidly enough in this scenario, but $Q_{\s
J}\sim 10^6$ could delay the evolution from deep resonance
sufficiently to prevent its destruction \cite{pea80}. Major problems
with this scenario were that (i) there was no mechanism for placing the
system in the resonances at the time of satellite formation and (ii) evolution
through the high-eccentricity phases could not be followed with the
analytic theory. Here, we describe a mechanism that addresses both of these
problems and adds support to Greenberg's hypothesis.

Constraints on the formation of the Galilean satellites of
Jupiter have been established by theoretical analysis and the Galileo
spacecraft observations.  The nearly coplanar and nearly circular
orbits of the four large satellites imply formation within a
dissipative, equatorial disk of gas and solid particles formed, perhaps, by
one of several processes during the formation of Jupiter 
\cite{pol91}.  Ganymede and Callisto are about
50\% ice and 50\% rock by mass \cite{sch86}. Europa
has a layer of water and ice 80 to 170 km thick \cite{and98b}.  The
large amount of relatively volatile 
material in Ganymede and Callisto implies that the disk temperature
remained sufficiently cool during the accretion process that icy
particles could persist.  Callisto may not be fully differentiated
\cite{and98a, and01}, a state that requires an accretion time scale
$\sim 10^6$ years to keep interior temperatures below 273 K 
\cite{ste86, ste01}.

Accretion models for the Galilean satellites have typically assumed an
initial disk with sufficient solid material to form the satellites and with
gas added to bring the mixture up to solar composition --- the so-called
minimum mass sub nebula (MMSN) \cite{pol91}. But a careful
evaluation of the processes in such a massive 
disk \cite{can02} indicate that temperatures
would generally be too high and accretion rates too fast to allow
formation of the satellites with their observed properties. In
particular, it would be difficult if not impossible to keep Callisto
only partially differentiated.  Therefore, it is likely that the
satellites formed near the end of Jupiter's formation when the flux 
of material into Jupiter through an accretion disk was greatly reduced
\cite{ste01}. Theoretical calculations \cite{war00} show that Jupiter
would limit this flux by opening a gap in the disk of material
surrounding the Sun when the planet's mass approached its current value.
Hydrodynamical simulations 
\cite{lub99,bry99,kle01,dan02} show that gas and particles
continue to trickle through the gap in two streams entering at the
inner and outer Lagrange points to form an accretion disk around
Jupiter. The surface mass density of the accretion disk resulting from this
scenario is sufficiently less than the MMSN to allow the slow accumulation
of the satellites in a cool environment.

By  modeling processes in this disk, Canup and Ward \cite{can02} have
found a scenario of satellite accretion that appears to satisfy all of
the observational and theoretical constraints imposed by the
satellites' current properties.  The important characteristics of the model,
for our purposes,
are that the formation time scales of the individual Galilean satellites are
similar and that the satellites migrate toward Jupiter while the disk
persists. The migration is the result of the gravitational interaction
of the satellites with spiral density waves generated in the disk
material at mean motion commensurabilities with the satellites 
(Lindblad and corotation resonances).  The satellites are not sufficiently
massive to open gaps in the disk and, thus, they execute the so-called Type I
drift \cite{war97}, whose time scale is inversely proportional to both the
disk surface mass density and the satellite mass. 

The key ingredient of the Canup-Ward model that allows an alternative
history of the Galilean satellite resonant system is that Ganymede's orbit
converges on those of Europa and Io because of its faster Type I
migration induced by its greater mass. Callisto is left behind
because of its smaller mass. Various choices of the
viscosity parameter $\alpha$ and opacity of the disk lead to many
different migration rates for the satellites, but all reasonable choices lead
to converging orbits of the inner three satellites [figs. 7c, 8c, and 9c
of \cite{can02}] because of 
Ganymede's more rapid migration. Any other model of formation will also
suffice for our purposes here, provided that the formation time scales
of all the satellites are comparable and that the disk persists long
enough for differential migration to lead to converging orbits \cite{leepeale}.

We demonstrate (Fig. 1) that the Laplace relation can be established
primordially during the final stages of satellite accretion by 
allowing Ganymede's orbit to migrate inward on the $10^5$ year time
scale of the Canup-Ward model \cite{can02}, whereas Europa and
Io initially migrate more slowly on a $2\times 10^5$ year time
scale. (The very rapid migration
rates implied for disk densities corresponding to the MMSN would most
likely preclude the capture of the satellites in any resonances during 
differential migration.)
The satellites start in an initial configuration where the orbits are
separated by distances greater than those appropriate to the 2:1 mean motion
commensurability.  The numerical calculations are done both by the
Wisdom-Holman \cite{wis91} symplectic integrator contained in the 
SWIFT package \cite{swift} and in a center-of-mass cartesian
coordinate system with a Bulirsch-Stoer algorithm.  The algorithms are
modified as described in \cite{lee02} to accommodate the migration and
eccentricity damping, and both algorithms yield similar
results. Europa is first captured into the  2:1 mean motion resonances
with Ganymede and Io is subsequently picked up in the 2:1 resonances with
Europa, simultaneously establishing the Laplace relation \cite{iorate}.  

The orbital eccentricities of the satellites are damped by interactions
with the gas disk [e.g., \cite{art93}] and 
eventually by tidal dissipation in the satellites [e.g.,
\cite{yod81}]. In the case of disk interactions, the ratio of the inverse time
scales is given by \cite{art93}
\begin{equation}
\frac{1/\tau_e}{1/\tau_a}=\frac{|\dot e/e|}{|\dot a/a|}=
9\frac{C_e}{C_a}\left(\frac{H}{a}\right)^{-2}\approx 30,  \label{eq:taue_taua} 
\end{equation}
where $H/a$ is the ratio of the scale height of the disk $H$
to the semimajor axis of the orbit $a$. $H/a\approx 0.1$ is assumed
constant in the disk, where the numerical value follows from 
the conditions in the Canup-Ward model \cite{can02}. 
$C_a\approx 3$ is the coefficient of the expression for $\dot
a/a$ \cite{art93,tan02}, and $C_e\approx 0.1$ is the coefficient 
in the expression for $\dot e/e$ \cite{art93}.  For the integrations, 
we assumed that the time scales for migration in $a$ and in eccentricity
damping are constant and, hence, the ratio of the inverse time
scales is fixed, with  $|\dot e/e|/|\dot a/a|=30$ from
Eq. 1 assumed for all three satellites.
The eccentricity damping causes the eccentricities to reach constant
asymptotic values (Fig. 1), which
are reduced if the ratio of the inverse time scales in
Eq. 1 is increased.  The evolution 
is reasonably independent of the rate of migration
up to a certain point. An example where the satellite migration rates
were 30 times larger led to a very similar evolution.

Although all of the resonance variables, including the
Laplace angle $\theta_5$, are librating in the primordial configuration
of the satellites (Fig. 1), the geometry of the system
differs from the current configuration, where all resonance variables,
except $\theta_4$, librate about either $0^\circ$ or $180^\circ$, as
described above. The  
centers of libration of $\theta_5$ and $\theta_2$ are near
$80^\circ$ instead of $180^\circ$ and that of $\theta_1$ is near
$-30^\circ$ instead of $0^\circ$ (Fig. 2). $\theta_4$
would normally librate 
about $180^\circ$ for smaller eccentricities, but here it is near
$-10^\circ$.  Only $\theta_3$ has the same libration center
near $0^\circ$ that it has now.  Conjunctions would no
longer occur near the lines of apsides of  
the orbits.  This is consistent with
the locus of stable libration centers for $\theta_5$ branching
symmetrically from $180^\circ$ as the system is driven deeper into
the resonances than the current configuration \cite{gre87}.  The
occurrence of libration centers of both $\theta_2$ 
and $\theta_5$ near $-80^\circ$ in some of our calculations is
verification of the symmetric branches. 

As the satellites move closer to Jupiter and the disk dissipates,
tidal dissipation in the satellites eventually dominates the
eccentricity damping. At the same time, the dissipation of tidal
energy in Jupiter from tides raised by the satellites results in the
torques that tend to push the satellites outward thereby tending to
increase the eccentricities [{\it e.g.}, \cite{yod81}].  Because
the current orbital eccentricities are smaller than the primordial
asymptotic values (Fig. 1), the tendency for the
eccentricities to decrease due to dissipation in the satellites
must dominate the effect of the tides raised on Jupiter that tend to
increase the eccentricities---at least at the end of migration. 
The tidal time scale for eccentricity damping is $|e/\dot e|\approx
6.5\times 10^4Q_{\s I}/f$ years for Io, which is derived from expressions
given in \cite{pcr80}.  Here $Q_{\s I} \approx 30\,{\rm to}\,100$
for rocky material and $f$ is a factor to
account for an inhomogeneous structure of Io, which can be as high as
13 for a two layer model consisting of a solid mantle over a molten
interior \cite{pea79}. The tidal time scale for eccentricity damping
is probably more than 6 times as long as the migration time
scale of the Canup-Ward model, and possibly 100 times as long as
the disk eccentricity damping time scale if
Eq. 1 is correct. This means that tides will not
take over the eccentricity damping until after the migration has
essentially stopped.   

After migration has stopped, we assumed that the dissipation in
Io is reducing its eccentricity on the same time scale as disk interaction
did during the migration ($\sim 6.7\times 10^3$ 
years) (Figs. 3 and 4). 
We maintained this time scale for the damping calculation to keep the
computer time within reasonable bounds. The algorithm is
such that forces are applied to Io that would cause the assumed
eccentricity damping if Io were isolated but, because it must drag down
the eccentricities of all the satellites locked in the resonances, the
actual time scale is much longer.  The
system naturally relaxes to the current configuration through a series
of states with offset libration centers, much like Greenberg
described \cite{gre87}. The eccentricities at the end of the calculation
have almost reached the values observed currently, and the libration centers
have migrated to the current geometry from their positions at the end
of migration (Fig. 2)---even to the detail
that $\theta_4$ has ended up circulating just as we observe now. 

The eccentricity of Ganymede, $e_3$, evolves in an
interesting way in conjunction with the behavior of $\theta_4$ 
(Figs. 3 and 4).
Initially, the center of libration of $\theta_4$ shifts from near $-10^\circ$
to near $180^\circ$ in about $10^4$ years while $e_3$ decreases.
At $1.5\times 10^4$ years, a sudden jump in $e_3$, perhaps from an
encounter with a secondary resonance \cite{tit89}, results in
$\theta _4$ changing to circulation (angle increasing or decreasing
without bound) 
where it remains until $t=5\times 
10^4$ years.  Here $e_3$ suddenly becomes smaller, perhaps from another
change in the effect of the secondary resonance, and $\theta_4$ again librates
about $180^\circ$.
After $\sim 10^4$ years, $\theta_4$ switches from libration to permanent
circulation due to the slow reduction in the
forced eccentricity while the free eccentricity remains near $0.0016$.
In another calculation with a different migration rate, the resonance variable
$\theta_4$  shows similar behavior but without the sudden changes in
$e_3$ and with different timing of events.
In both cases, $\theta_4$ ends up circulating with $e_3 \approx 0.0016$.
This is about the current value of Ganymede's free
eccentricity, but there may be circumstances where less
excitation of the free eccentricity leaves $\theta_4$ librating.

A primordial origin of the Laplace relation relieves Jupiter of
providing sufficient tidal torque over history to expand the orbits
enough for the completely tidal origin.  
However, it has been argued that even if the origin of the resonances
is primordial, the  average $Q_{\s J}$ cannot be far from 
the bounds $6\times 10^4\ltwid Q_{\s J}\ltwid 2\times 10^6$ established by
\cite{gol66,yod79,pea80,yod81,gre87}. The current dissipation in 
Jupiter must still be sufficient to prevent the destruction
of the resonances from the tidal dissipation in Io and might
be sufficiently high to preserve the system in equilibrium
\cite{pea80,yod81}. In contrast, could it be  
that we are witnessing the further decay of the Laplace 
resonance due to the dissipation in Io \cite{gre82,gre87}?  Because the energy
dissipated must come from the orbits, a continuing decay of the
Laplace configuration would result in an increasing $n_1$.
Measures of $\dot n_1=\ddot\lambda_1$, based on 300 years of precise
ephemerides of the Galilean satellites coupled with current precise
observations, imply either an acceleration of Io's mean motion,
indicating further decay of the resonant configuration \cite{gol95,aks01}, or
insufficient deceleration \cite{lie87} from Jupiter torques on Io to
accommodate the high heat flux from Io in an equilibrium
eccentricity configuration.  The heat flux is now estimated to be 
about $3\,{\rm W/m^2}$ \cite{vee02}. Although the
measurements of $\dot n_1$ remain controversial, we should keep in
mind that a heat flux from Io of $3\,{\rm W/m^2}$ would require a
current $Q_{\s J}\approx 2\times 10^4$ to maintain the system in
equilibrium, which is a factor of 3 lower than the minimum historical
average $Q_{\s J}=6\times 10^4$ from the proximity of Io to Jupiter
\cite{gol66}.   But there is no reason to believe that the current
$Q_{\s J}$ should be constrained much by a bound on its historical
average \cite{pea99}.   

A primordial origin of the Laplace relation among the Galilean
satellites, created by differential migration of the satellites in the
gaseous disk, with the help of tidal dissipation, reproduces every
detail of the current configuration.
Changing assumptions will change details of this
scenario such as the asymptotic values of the eccentricities during
the migration; a different damping rate may change the circulation
of $\theta_4$.  But as long as the satellites exist simultaneously in
the disk and the more rapid migration of Ganymede forces the inner-three
orbits to converge and capture into the Laplace resonance
results \cite{otherres}, the qualitative features of this alternative
early history will persist. 

\bibliography{preprint}

\begin{thebibliography}{10}
\bibitem{free}
Free and forced eccentricities are often referred to in discussions of
  eccentricity type orbital resonances at commensurabilities of the mean
  motions. In a lowest order eccentricity type resonance between two
  satellites, where for example the resonance variable $\lambda_1-2\lambda_2+
  \varpi_1$ is librating about $0^\circ$ ($n_1-2n_2+\dot\varpi_1=0$ on
  average), conjunctions must occur when satellite 1 is near its periapse for
  stability. This imposes the constraint $\langle \dot\varpi_1\rangle=-\langle
  n_1-2n_2 \rangle$, where the angled brackets denote averaged values. Because
  $\langle \dot\varpi_1\rangle$ depends on the magnitude of the eccentricity
  $e_1$, the resonant interaction will force $e_1$ to the appropriate value to
  accommodate the resonance. If there is a nonzero libration amplitude, the
  eccentricity will fluctuate about the forced mean value. The magnitude of the
  fluctuation in $e_1$ is called the free eccentricity, which can be damped to
  zero.
\bibitem{leepeale}
Lee and Peale \cite{lee02} have shown that capture into the resonances at the
  2:1 commensurability during differential migration of the two planets in
  coplanar orbits about the star GJ 876 occurs easily for a reasonably wide
  range of initial conditions. Capture and subsequent configuration are
  independent of the migration rate unless that rate is unreasonably large, and
  the capture and configuration are independent of the functional dependence of
  the migration on orbital semimajor axes. The only thing that matters is that
  the orbits converge.
\bibitem{iorate}
We have arbitrarily assigned the full masses to the satellites and assumed that
  Io and Europa migrate on the same time scale to demonstrate the hypothesis,
  whereas migration will ensue as the satellites are accreting and Io will
  generally migrate faster than Europa once its mass exceeds that of the
  latter. The necessary condition here for the success of the establishment of
  the Laplace relation is that Io migrate more slowly than Europa after Europa
  is captured into resonance with Ganymede, which capture accelerates Europa's
  migration. The migration rate history thus depends on the surface mass
  density of the disk and the accretion history determining the satellite
  masses as a function of time. There is enough flexibility in the disk and
  accretion model to effect the necessary relative rates.
\bibitem{otherres}
The establishment of the Laplace relation is not a certain outcome for
  differential migration of the satellites, although some type of resonance
  configuration does seem certain. Minor changes in the initial conditions,
  migration rates, or eccentricity damping rates sometimes led to other
  configurations. For example, in one case the system ended up with
  Ganymede-Europa in 2:1 resonances and Europa-Io in 7:3 resonances. Accretion
  of the satellites may not have been complete at the time of capture into the
  resonances, but two identical calculations---one with the full satellite
  masses and the other with half these masses---led to very similar Laplace
  configurations. Even if the mass ratios vary during migration, the
  configuration at the end of migration should be the same. The conditions for
  the various end configurations and an estimate of the probability of the
  system selecting the Laplace configuration will be subjects of a future
  paper.
\end{thebibliography}

\bibliographystyle{Science}

\begin{scilastnote}
\item We thank R. Canup and W. Ward for
sending us an early preprint of their paper on Galilean satellite
origin. This research was supported in part by NASA PGG grants NAG5
3646 and NAG5 11666 and SSO grants NAG5 7177 and NAG5 12087.
\end{scilastnote}

\clearpage

\begin{figure}
\epsfxsize=5.5truein \epsfbox{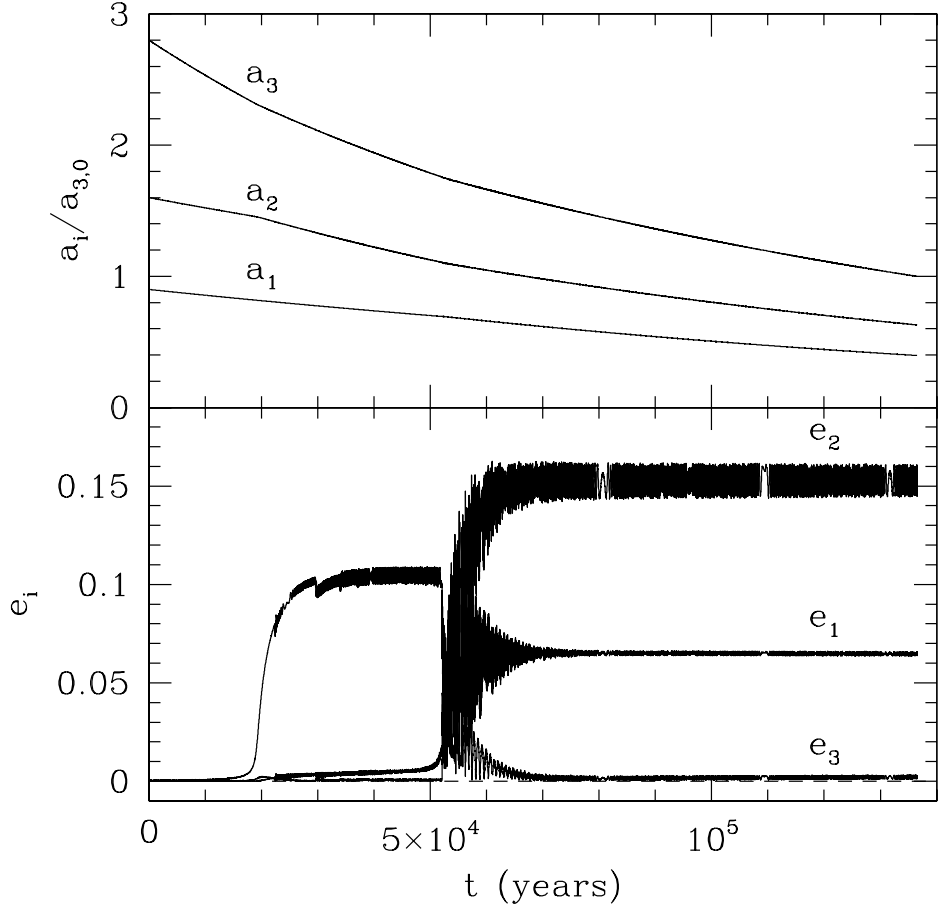}
\caption{Nebula-induced evolution of Galilean satellites into
the Laplace resonance.  The semimajor axes are normalized by Ganymede's
current distance from Jupiter $a_{3,0}$. The imposed inward migration
and eccentricity damping time scales are $|a_3/\dot
a_3|\approx 10^5$ years, $|a_1/\dot a_1|=|a_2/\dot a_2| \approx 2\times
10^5$ years, and  $|e_i/\dot e_i| = |a_i/(30\dot a_i)|$
$i=1$, 2, and 3. Circular coplanar orbits are assumed initially.}
\end{figure}

\begin{figure}
\epsfxsize=5.5truein \epsfbox{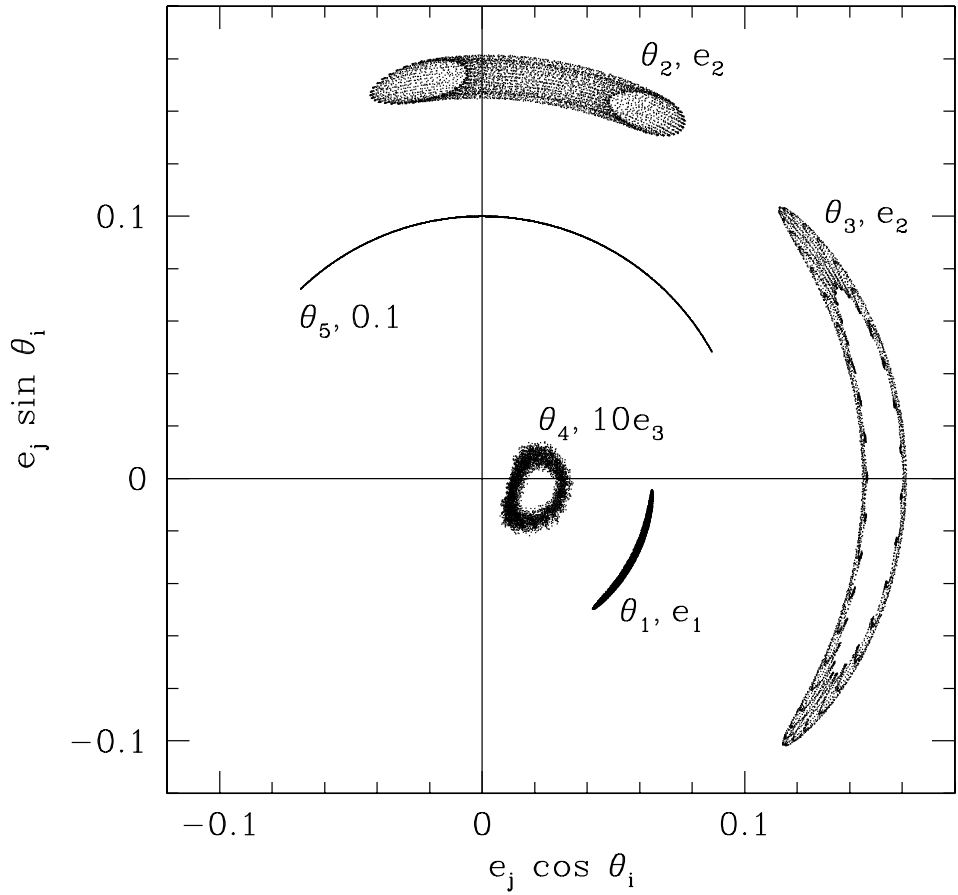}
\caption{Libration of 2:1 mean motion resonance variables and
the Laplace angle for the asymptotic state of the resonant Galilean
satellite configuration shown in Fig. 1. The
$\theta_i$ are defined in the text. The coefficients $e_j$
corresponding to the $\theta_i$ are indicated in the figure.
Librations are shown for 250 years with 
all migration and damping removed.}
\end{figure}

\begin{figure}
\epsfxsize=5.5truein \epsfbox{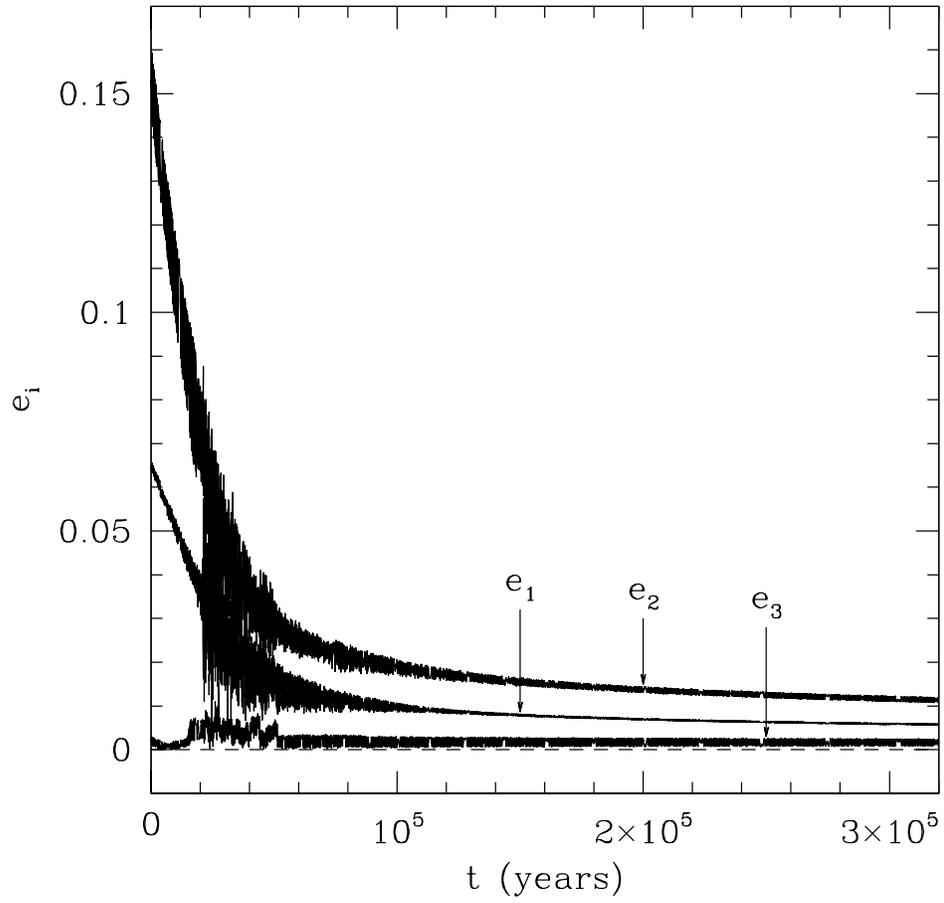}
\caption{Relaxation of Galilean satellite system to the current
configuration with migration halted and with eccentricity damping in
Io alone to simulate the effects of tidal dissipation.}
\end{figure}

\begin{figure}
\epsfxsize=5.5truein \epsfbox{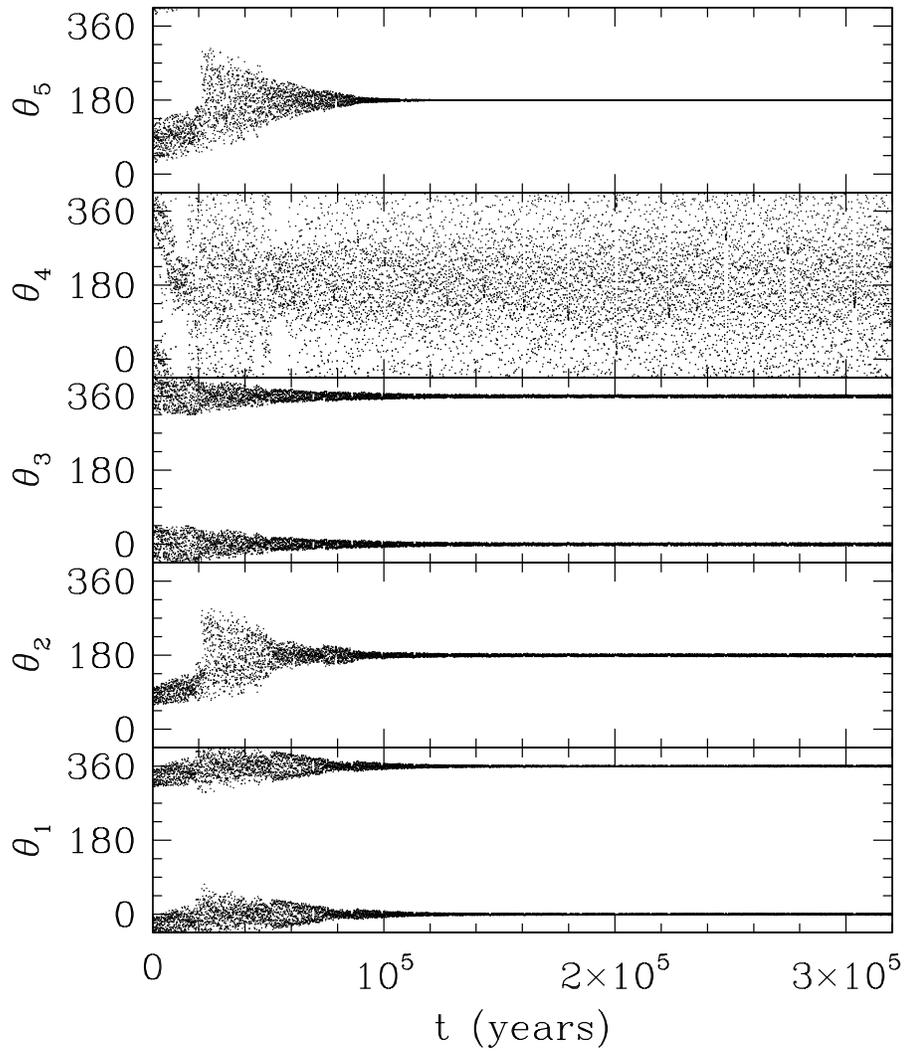}
\caption{Behavior of resonance variables for the evolution
described in Fig. 3.}
\end{figure}

\end{document}